\documentstyle[12pt]{article}
\newcommand{\beqn}{\begin{eqnarray}}
\newcommand{\eeqn}{\end{eqnarray}}
\newcommand{\be}{\begin{equation}}
\newcommand{\ee}{\end{equation}}
\newcommand{\ba}{\begin{array}}
\newcommand{\ea}{\end{array}}
\newcommand{\R}{{\rm\bf R}}
\newcommand{\C}{{\rm\bf C}}

\newcommand{\re}{\ref}
\newcommand{\ci}{\cite}
\newcommand{\la}{\label}
\newcommand{\bfr}{\begin{flushright}}
\newcommand{\efr}{\end{flushright}}
\newcommand{\bfl}{\begin{flushleft}}
\newcommand{\efl}{\end{flushleft}}

\newcommand{\ga}{\gamma}

\newcommand{\ds}{\displaystyle}

\newcommand{\om}{\omega}

\newcommand{\lam}{\lambda}

\newcommand{\br}{|\kern-.25em|\kern-.25em|}

\begin{document}
 \renewcommand{\theequation}{\thesection.\arabic{equation}}
\newtheorem{theorem}{Theorem}[section]
\renewcommand{\thetheorem}{\arabic{section}.\arabic{theorem}}
\newtheorem{definition}[theorem]{Definition}
\newtheorem{deflem}[theorem]{Definition and Lemma}
\newtheorem{lemma}[theorem]{Lemma}
\newtheorem{example}[theorem]{Example}
\newtheorem{remark}[theorem]{Remark}
\newtheorem{remarks}[theorem]{Remarks}
\newtheorem{cor}[theorem]{Corollary}
\newtheorem{pro}[theorem]{Proposition}
\mathsurround=2pt
\newcommand{\HH}{{\rm\bf H}}
\newcommand{\pr}{\prime}
\def\N{{\rm I\kern-.1567em N}}                              
\def\No{\N_0}                                               
\def\R{{\rm I\kern-.1567em R}}                              
\def\C{{\rm C\kern-4.7pt                                    
\vrule height 7.7pt width 0.4pt depth -0.5pt \phantom {.}}\,}
\def\Z{{\sf Z\kern-4.5pt Z}}                                
\def\n#1{\vert #1 \vert}                                    
\def\nn#1{\Vert #1 \Vert}                                   
\def\Re {{\rm Re\, }}                                       
\def\Im {{\rm Im\,}}                                        
\newcommand{\loc}{\scriptsize loc}
\newcommand{\const}{\mathop{\rm const}\nolimits}
\newcommand{\supp}{\mathop{\rm supp}\nolimits}
\newcommand{\mod}{\mathop{\rm mod}\nolimits}
\newcommand{\ow}{\overrightarrow}

\begin{titlepage}
\hspace{6cm} 
{\it Russ. J. Math. Physics} {\bf 9} (2002), no. 2, 153-160  
~\bigskip\bigskip\\
 \begin{center}
{\Large\bf
Energy-momentum relation for solitary waves
\medskip\\
of relativistic wave equations}\\
 \vspace{1cm}
{\large T.V. Dudnikova
\footnote{Supported partly by research grants of 
DFG (436 RUS 113/615/0-1),
INTAS-OPEN-97-10812 and RFBR (01-01-04002).}
}\\
{\small\it Mathematics Department\\
Elektrostal Polytechnic Institute\\
Elektrostal, 144000 Russia\\
e-mail:~~ dudnik@elsite.ru}
\bigskip\\
{\large A.I. Komech
\footnote{Supported partly
by the Institute of Mathematics of the Vienna University, 
by Max-Planck Institute for Mathematics in the Sciences
(Leipzig) and by the START project (FWF Y137-TEC)
of N.J. Mauser.}
}\\
{\small\it Mechanics and Mathematics Department \\ Moscow State University\\
Moscow, 119899 Russia\\
e-mail:~~ komech@mech.math.msu.su}
\bigskip\\
 {\large H. Spohn}\\
{\small\it Zentrum Mathematik\\
Technische Universit\"at\\
M\"unchen D-80290, Germany\\
 e-mail:~spohn@ma.tum.de}\\
 \vspace{1.2cm}
 \end{center}

\begin{abstract}
Solitary waves of relativistic invariant nonlinear wave 
equation with symmetry group $U(1)$  are considered.
We prove that the energy-momentum relation for
spherically symmetric solitary waves coincides with the
Einstein  energy-momentum relation for point particles.
\end{abstract}
\end{titlepage}
\section{Introduction}

The paper concerns the old problem of mathematically describing elementary particles in  field theory.
The problem was raised in classical electrodynamics
after the discovery of the electron by Thomson in 1897.
Abraham  realized that  a point electron would be
unstable due to  infinite self-energy,
and introduced the model of the ``extended''
electron with  finite electrostatical energy.
However, the electrostatic model is also unstable
 due to electrostatic repulsion.
The corresponding tension of the extended electron
 was analyzed by Poincar\'e in \ci{P}.
To avoid the instability problem for point particles,
Einstein suggested that particles could be described 
as singularities
of  solutions to the field equations  \ci{Ein}.
To obtain stationary finite energy solutions, 
Born and Infeld introduced a ``unitary field'', which is a
nonlinear modification of the Maxwell equations \ci{B,BI}.
The last approach was not developed further because 
the relation to 
the corresponding quantum version was not clarified.

Rosen \ci{Rn} was the first who proposed  
a description of particles for the coupled Klein-Gordon-Maxwell equations,
which are invariant with respect to the Lorentz group and 
to the global gauge group $U(1)$:
the particle at rest is described by
a finite energy solution that has ``Schr\"odinger's'' form
$\psi(x)e^{i\om t}$ (``nonlinear eigenfunctions'' or ``solitary waves'').
The  particle with the nonzero velocity $v$,  $|v|<1$,
is obtained by the corresponding Lorentz (or Poincar\'e) transformation.
Some numerical analysis has been done,
 showing the existence of radial solutary waves for a finite range 
of $\om$. The particle in the ``normal'' state is identified with
the solitary wave of minimal energy.

The existence of radial solitary waves (``ground states'')
 and nonradial solitary waves
(``excited states'' with nonzero angular momentum) 
has been analyzed numerically 
by many authors for diverse Lagrangian field theories
 \ci{FFK,FLR}, \ci{Rn}-\ci{Sol}, \ci{W}.
For the complex scalar field,
most general results were achieved by Beresticky and 
Lions \ci{BL}. For the nonlinear Dirac equation, the existence is proved 
in \ci{BCDM, ES}. 
For the Maxwell-Dirac field, the existence was proved by
Esteban, Georgiev and S\'er\'e in \ci{EGS}.

The next step after establishing the existence of 
solitons 
would be to study their stability as
one of the main features of elementary particles.
A nonrigourous analysis in \ci{Rn} suggests an instability 
of solitons with  negative total energy treated as 
a negative mass of the soliton.
The first rigourous proof of instability 
of solitons with $\om =0$ was given by
Derrick in \ci{Der}. 
The proof also concerns negative energy and 
is based on the virial Derrick-Pokhozhaev
identity \ci{Der,Po}.

These instability results hindered 
the development of the approach to elementary particles 
as the solitons. On the other hand, 
the solitons with $\om\ne 0$
can be stable; this follows from the general criterion 
discovered by Grillakis, Shatah and Strauss and other
authors (see \ci{GSS} and the references therein).
Note that in \ci{BP}
the asymptotic stability of solitons 
was proved  for the 1D nonlinear Schr\"odinger equation
 when $\om\ne 0$.

The key role of elementary particles suggests that the set
of all moving solitons forms a global attractor for all
finite energy solutions to the dynamical field equations.
However, this is still an unproved conjecture.

Finally, it would be  of importance to 
develop a particle-like dynamics for moving solitons.
We make a step in this direction
for relativistic-invariant scalar Klein-Gordon equations.
Namely, we  prove that,
for spherically symmetric solitary waves (constructed in \ci{BL}),
the energy-momentum relation coincides with that of
a relativistic particle.

Furthermore, we suggest that
this fact fails for non-spherically symmetric solitary 
waves
of some angular structure if such solitary waves
exist indeed.

A further step would be a justification of
effective dynamics for a solitary wave in an external
slowly varying potential. A result of this type
was obtained in \ci{KKS} for the solitons of the Abraham model of classical electrodynamics and 
in \ci{Sp}  for localized wave packets
of the  Dirac equation.
Recently, an effective dynamics was announced for the solitons 
of the nonrelativistic nonlinear Hartree equation \ci{FTY}.

\section{Standing solitary waves}
Consider the relativistic-invariant nonlinear wave equation
\be\la{1}
\ddot \psi(x,t)=\triangle \psi(x,t)+f(\psi(x,t)),\quad x\in\R^n,
\ee
where $n\ge 1$ and  $\psi\in\C^d$, $d\ge 1$.
Assume that
$$
f(\psi)=-\nabla_\psi V(\psi),\quad \psi\in \C^d,
$$
where $V$ is a real potential, and
$\nabla_{\psi}$ stands for the gradient
with respect to $u=\Re \psi$ and $v=\Im\psi$;
in other words, $\nabla_\psi=\nabla_u+i\nabla_v$.
Then Eq. (\ref{1}) formally becomes a Hamiltonian system with the  Hamiltonian functional
\be\la{H}
H(\psi,\dot \psi)=
\int\left[
\frac{|\dot \psi|^2}{2}+\frac{|\nabla \psi|^2}{2}+V(\psi)
\right]\,dx.
\ee
Further, assume that
$V(\psi)={\cal V}(|\psi|)$. Then
Eq. (\ref{1}) is $U(1)$-invariant, i.e.,
$$
~~f(e^{i\theta}\psi)=e^{i\theta}f(\psi),\quad\theta\in\R.
$$
Consider a (standing) solitary wave
$\psi_0(x,t)=a(x)e^{-i\omega t}$ with the energy
\be\la{E0}
H(\psi_0,\dot \psi_0)=
\int\left(
\frac{\omega^2 |a|^2}{2}+\frac{|\nabla a|^2}{2}+{\cal V}(|a|)
\right)\,dx.
\ee
The amplitude $a(x)$ is a solution of the  Helmholtz stationary
nonlinear equation
\be\la{a}
-\omega^2 a(x)=\triangle a(x)+f(a(x)),\quad x\in\R^n.
\ee
The existence of nonzero solitary waves was 
proved in \ci{BL}
under the following assumptions:
$$
\ba{ll}
{\bf S0}
\quad\quad \quad\quad\quad
& d=1,\quad f\in C(\C),\quad  f(0)=0,
\quad\quad \quad\quad\quad
\\
~\\
{\bf S1}
\quad\quad \quad\quad\quad
&  -\infty<
\ds\lim\limits_{a\to 0+}{f(a)}/a+\om^2<0,
\quad\quad \quad\quad\quad
\\
~\\
{\bf S2}
\quad\quad \quad\quad\quad
&
  \exists a_0>0:
{\cal V}(a_0)-\ds\frac{\omega^2a_0^2}{2}<0,
\quad\quad \quad\quad\quad
\\
{\bf S3}
\quad\quad \quad\quad\quad
&\exists\alpha\ge 0:
 f(a)=-\alpha a^l+o(a^l),\,\,a\to\infty,\,\,\,\,\,\,\,
l:=\ds\frac{n+2}{n-2}.
\quad\quad \quad\quad\quad
\ea
$$
Further, the solution $a(x)$ to (\ref{a}) is real and
 satisfies the following properties:
$$
\ba{ll}
{\bf A1}\quad\quad\quad\quad\quad\quad\quad   &a(x)=R(|x|),
\quad\quad\quad\quad\quad\quad\quad\quad\quad\quad\\
~\\
{\bf A2}\quad\quad\quad\quad\quad\quad\quad \exists C,\delta >0:&
|R^{(k)}(r)|\le C e^{-\delta r},\,\,\,\,k=0,1,2.
\quad\quad\quad\quad\quad\quad\quad\quad\quad\quad
\ea
$$
Property {\bf A1}
means that the solitary wave thus constructed  is
a radial (spherically symmetric)  ``ground state''.
Below we discuss  the excited states
with higher angular momentum for $n=2$.
\section{Moving solitary waves}

Consider a (moving) solitary wave with velocity $v\in\R^n$:
$$
\psi_v(x,t)=\psi_0(\Lambda_v (x,t)).
$$
Here $|v|<1$ and $\Lambda_v$ is a Lorentz 
transformation  with  velocity $v$:
$$
\Lambda_v(x,t)=(\gamma(x^\Vert-vt)+x^\perp, \gamma(t-v x)),
$$
where $x^\Vert+x^\perp=x$, $x^\Vert\Vert v,$
$x^\perp \perp v,$
$ \gamma=\ds\frac{1}{\sqrt{1-v^2}}.$
In other words,
$$
\psi_v(x,t)=\psi_0(\gamma(x^\Vert-vt)+x^\perp,\gamma(t-vx))
=a(\gamma(x^\Vert-vt)+x^\perp)e^{-i\omega \gamma(t-vx)}.
$$
Note that
 $\psi_v(x,t)$ is a solution of (\ref{1}).

\setcounter{equation}{0}
\section{Energy-momentum relation for radial states}
For $v\in \R^n$, $|v|<1$, we denote by
$$
E_v:=H(\psi_v,\dot \psi_v),\,\,\,
P_v:=-\Re\int \dot \psi_v\overline{\nabla \psi_v}\,dx
$$
the energy and momentum of
the moving solitary wave, respectively.
Let us assume in what follows that
$$
{\bf S4}\quad\quad\quad\quad\quad\quad\quad
\quad\quad\quad\quad\quad\quad\quad\quad\quad
\,\,\,{\cal V}(a)+\ds\frac{\omega^2a^2}{2}\ge 0,~~a\ge 0.
\quad\quad\quad\quad\quad\quad\quad\quad\quad
\quad\quad\quad\quad\quad\quad\quad\quad\quad
$$

In this section, we consider spherically symmetric nonzero
solitary waves (i.e. waves with
zero angular momentum). This spherical symmetry is a
typical property of a ``ground state'' with minimal energy.
\begin{theorem}\la{Th}
Let  $a(x)e^{-i\om t}$ be a standing
solitary wave for (\re{1}),  and let {\bf A1}, {\bf A2} 
hold. Then  the following ``particle-like'' 
energy-momentum relation  holds:
\be\la{A}
E_v=\frac{E_0}{\sqrt{1-v^2}},\,\,\,
P_v=\frac{E_0v}{\sqrt{1-v^2}}.
\ee
Here $E_0>0$ for $\om\ne 0$.
\end{theorem}
{\bf Remark}
For a nonzero solitary wave, we have
 $E_0:=H(\psi_0,\dot \psi_0)> 0$.
In fact, if $H(\psi_0,\dot \psi_0)=0$, then
 $a(x)\equiv$const
by {\bf S4} and (\ref{E0}).
However, this implies $a(x)\equiv 0$ by {\bf A2}.
\medskip\\
{\bf Proof of Theorem 4.1.}
{\it Step 1.}
We choose the coordinates in such a way that
 $v=(|v|,0,\dots,0)$.
Below we write everywhere $v$ instead $|v|$
to simplify the notations.
Let us write
$$
y_1=\gamma(x_1-vt),\,\,y_k=x_k;\,\,\,y=(y_1,...,y_n).
$$
Then
\beqn
\psi_v(x,t)&=& a(y) e^{-i\omega\gamma(t-vx_1)},\nonumber\\
\dot \psi_v(x,t)&=&\left(-\gamma v
 (\nabla_1 a)(y)
-i\gamma\omega a(y)\right)e^{-i\omega\gamma(t-vx_1)},
\nonumber\\
\nabla_1 \psi_v( x,t)&=&
\left(\gamma (\nabla_1 a)(y)
+i\gamma\omega v a(y)\right)e^{-i\omega\gamma(t-vx_1)},
\nonumber\\
\nabla_k \psi_v(x,t)&=&
(\nabla_k a)(y) e^{-i\omega\gamma(t-vx_1)},\quad k=2,...,n.
\nonumber
\eeqn
Substituting these expressions into $E_v$, we obtain
\beqn
E_v&=&\int\left[
\frac{\dot \psi_v^2}{2}+\frac{|\nabla \psi_v|^2}{2}+V(|\psi_v|)
\right]\,dx
\nonumber\\
&=&\int\left(\frac{1}{2}\left[
|\nabla_1a(y)|^2\gamma^2( v^2+1)
+|a(y)|^2\omega^2\gamma^2(v^2+1)
+\sum\limits_{k=2}^{n}
|\nabla_k a(y)|^2\right]+V(|a(y)|)\right)\,dx.\nonumber
\eeqn
Then
$$
E_v=\int\left(
\frac{1}{2} \sum\limits_{k=2}^{n}
|\nabla_k a|^2 +
\frac{\gamma^2}{2}(v^2+1)\left[
|\nabla_1 a|^2 +|a|^2\omega^2\right]
+V(|a|)
\right)\frac{1}{\gamma}\,dy.
$$
Write
\beqn
&&I_0=\ds\frac{1}{2}\int |a(y)|^2\,dy,
\quad V_0=\int V(|a(y)|)\,dy,\la{IVA}\\
~\nonumber\\
&&I_k=\ds\frac{1}{2}\int |\nabla_k a(y)|^2\,dy,\quad k=1,...,n.
\la{IVB}
\eeqn
Hence, in the notation (\ref{IVA}) and (\ref{IVB}),
 we represent $E_v$ as
\beqn\la{ev}
E_v&=&\frac{1}{\gamma}
\left(\sum\limits_{k=2}^{n} I_k
+\gamma^2(v^2+1)(I_1+I_0\omega^2)+V_0\right),\\
\la{e0}
E_0&=&\sum\limits_{k=1}^{n} I_k
+I_0\omega^2+V_0.
\eeqn
{\it Step 2.} We now derive
(\ref{A}) from this representation for $E_v$
with the help of the following lemma. 
\begin{lemma}\la{l1} (\ci{BL, Der, Po})
The following Derrick-Pokhozhaev identity holds:
\be\la{Pi}
-(n-2)
 \frac{1}{2}\int
 |\nabla a(y)|^2\,dy=
n\int \left[
V(|a(y)|)-\frac{1}{2}\omega^2 a^2(y)
\right]\,dy.
\ee
\end{lemma}
{\bf Proof.} Let us explain the proof formally. Relation
(\ref{a}) implies the variational identity
$$
-\delta\frac{1}{2}\int|\nabla a|^2\,dx=
\delta\int\left[
V(|a|)-\frac{1}{2}\omega^2 a^2
\right]\,dx.
$$
Therefore,
$$
-\frac{d}{d\sigma}\Big|_{\sigma=1}
\frac{1}{2}\int\Big|\nabla_x a\Big(\frac{x}{\sigma}\Big)\Big|^2\,dx=
\frac{d}{d\sigma}\Big|_{\sigma=1}
\int\left[
V\Big(a\Big(\frac{x}{\sigma}\Big)\Big)-\frac{1}{2}\omega^2 a^2\Big(\frac{x}{\sigma}\Big)
\right]\,dx.
$$
Equivalently,
$$
-\frac{d}{d\sigma}\Big|_{\sigma=1}
\sigma^{n-2} \frac{1}{2}\int
 |\nabla_y a(y)|^2\,dy=
\frac{d}{d\sigma}\Big|_{\sigma=1}
\sigma^{n}\int \left[
V(a(y))-\frac{1}{2}\omega^2 a^2(y)
\right]\,dy.
$$
This gives the Pokhozhaev identity (\ref{Pi}),
or (in the notation of (\ref{IVA}) and (\ref{IVB}))
\be\la{5}
-(n-2)\sum\limits_{k=1}^{n} I_k=
n\left[
V_0-\omega^2 I_0\right].
~~~~~
\Box
\ee
{\it Step 3.}
To derive  (\ref{A}), let us eliminate $V_0$ and $\om^2$
from  (\ref{ev}) and (\ref{5}) obtaining
\be\la{omegaI0}
\omega^2 I_0=\frac{n-2}{n}(I_1+...+I_n)+V_0.
\ee
Hence,  $E_0$ and $E_v$ (see (\ref{ev}) and (\ref{e0}))
become
\beqn\la{e00}
E_0\!\!&=&\!\!\frac{2n-2}{n}\sum\limits_{k=1}^nI_k+2V_0,\\
\la{eve}
E_v\!\!&=&\!\!
\frac{1}{\gamma}\left[\sum\limits_{k=2}^{n} I_k
+\gamma^2(v^2+1)\Big(I_1+
\frac{n-2}{n}\sum\limits_{k=1}^nI_k+V_0\Big)+
V_0 \right]\nonumber\\
&=&\!\!\!
\frac{1}{\gamma}
\left[I_1\gamma^2(v^2\!+\!1)\Big(1\!+\!\frac{n-2}{n}\Big)
+\sum\limits_{k=2}^nI_k
\Big(1\!+\!\gamma^2(v^2+1)\frac{n-2}{n}\Big)
\!+\!
V_0(\gamma^2(v^2\!+\!1)\!+\!1)
\right].\,\,\,\,\,\,\,
\eeqn
Note that $\gamma^2(v^2+1)+1=2\ga^2$
and
$\ds 1+\gamma^2(v^2+1)\frac{n-2}{n}=
\ga^2\frac{2n-2}{n}-\ga^2\frac{2v^2}{n}$. 
Therefore, by using (\ref{e00}), we can represent
$E_v$  as
\beqn\la{ever}
E_v&=&
\ga\frac{2n-2}{n}\sum\limits_{k=1}^nI_k+
\ga\frac{2v^2}{n}\Big(I_1(n-1)-\sum\limits_{k=2}^nI_k\Big)
+\ga2V_0\nonumber\\
&=&
\gamma E_0+\gamma\frac{2v^2}{n}\Big(I_1(n-1)-
\sum\limits_{k=2}^nI_k\Big).
\eeqn
Using assumption {\bf A1},
we obtain 
$I_1=I_2=...=I_n$, which implies
 identity (\ref{A}) for $E_v$.\\
~\\
{\it Step 4.} We have
\beqn
P_v&=&
-\Re\int \dot \psi_v\overline{\nabla \psi_v}\,dx
\nonumber\\
&=&
-\Re\int(-\gamma v(\nabla_1 a)(y)-i\omega\gamma a(y))
\Big(
\overline{\gamma (\nabla_1 a)(y)+i\omega\gamma va(y)},
\overline{\nabla_2 a(y)},\dots,
\overline{\nabla_n a(y)}
\Big)\frac{1}{\gamma}\,dy.
\nonumber
\eeqn
Since $a(x)$ is a real function, we obtain (in the
notation of (\ref{IVA}) and (\ref{IVB}))
\beqn\la{6}
P_v=\Big(\gamma v\int(|\nabla_1 a(y)|^2+\omega^2 a^2(y))
\,dy,0,\dots,0\Big)
=\Big(\gamma v 2(I_1+\omega^2 I_0),0,\dots,0\Big),
\eeqn
because $\nabla_1 a(y)$ is an even function
with respect to any variable $y_2,...,y_n$,
while $\nabla_k a(y)$ is an odd function of $y_k$, $k=2,...,n$.
Using (\ref{omegaI0}) and (\ref{e0}),
we obtain
\be\la{7}
\gamma v 2(I_1+\omega^2 I_0)=
\gamma v E_0+\frac{2\gamma v}{n}
\Big((n-1)I_1-\sum\limits_{k=2}^n I_k\Big).
\ee
Therefore, relations (\ref{6}) and (\ref{7}),
together with $I_1=I_2=...=I_n$, imply
$$
P_v=(\ga v E_0,0,\dots,0),
$$
 what yields (\ref{A}) for $P_v$.
\hfill$\Box$
\medskip\\
{\bf Remarks.}
i) Formulas (\ref{ever}) and (\ref{omegaI0}) 
show that identity  (\ref{A}) holds for general 
non-radial solitary waves if and only if
\be\la{add}
I_2+...+I_n=I_1(n-1).
\ee

ii) Relation (\ref{A}) implies
the Einstein identity $E_0=m_0$ with $m_0>0$ (see the remark after Theorem
 4.1).

iii) For $\omega =0$,
condition {\bf S4} contradicts {\bf S2}.
Hence, in this case we have $E_0\le 0$, i.e.,
 ``the mass'' is negative.
This corresponds to the instability of  solitons 
with $\om=0$, what was proved in \ci{Der}.
This fact forces us to consider solitary waves
with $\omega\not=0$ for $U(1)$-invariant equations.

\setcounter{equation}{0}
\section{Nonradial excited states}
In the previous section we have considered
a spherically symmetric solitary wave (``ground state'') 
with zero angular momentum.
For $n=1$, the condition (\ref{add}) obviously holds 
(and condition (\ref{A}) as well) for all 
solitary  waves.
Let us prove that (\ref{A}) holds for $n=2$ for non-radial
solitary  waves that describe the  excited states with
nonzero angular momentum.
Introduce polar coordinates by setting
$x_1=r\cos\varphi$ and $x_2=r\sin\varphi$.
\begin{lemma}
Let  $\psi_0(x,t)=a(x)e^{-i\omega t}$
be a solitary wave for (\ref{1}) with
 $a(x):=R(r)e^{ik\varphi}$,
where $k$ is  an integer, $R(r)$ is real valued,
$R(0)=0$, and {\bf A2} holds.
Then relation (\ref{A}) is satisfied.
\end{lemma}
{\bf Proof.} In the polar coordinates
$$
\nabla_1 a=R'(r)\cos k\varphi
-ik\ds \frac{R(r)}{r}\sin k\varphi,\,\,\,\,
\nabla_2 a=R'(r)\sin k\varphi  +
ik\ds \frac{R(r)}{r}\cos k\varphi.
$$
Therefore,
\beqn
I_1&\equiv&\frac{1}{2}\int|\nabla_1 a|^2\,dy_1d y_2=
\frac{1}{2}
\int\limits_{0}^{\infty}rdr\int\limits_{0}^{2\pi}d\varphi
\Big[|R'(r)|^2\cos^2 k\varphi  +k^2
\ds \frac{|R(r)|^2}{r^2}\sin^2 k\varphi\Big]\nonumber\\
&=&
\frac{\pi}{2}
\int\limits_{0}^{\infty}
\Big[|R'(r)|^2  +k^2
 \frac{|R(r)|^2}{r^2}\Big]\,r\,dr,\nonumber\\
I_2&\equiv&\frac{1}{2}\int|\nabla_2 a|^2\,dy_1d y_2=
\frac{1}{2}
\int\limits_{0}^{\infty}rdr\int\limits_{0}^{2\pi}d\varphi
\Big[|R'(r)|^3\sin^2 k\varphi  +k^2
\ds \frac{|R(r)|^2}{r^2}\cos^2 k\varphi\Big]\nonumber \\
&=&
\frac{\pi}{2}
\int\limits_{0}^{\infty}
\Big[|R'(r)|^2  +k^2
\ds \frac{|R(r)|^2}{r^2}\Big]\,r\,dr.\nonumber
\eeqn
Hence, $I_1=I_2$, i.e., (\ref{add}) holds, which implies  (\ref{A}) for $E_v$.
$\hfill\Box$\\

Now let us  prove identity (\ref{A}) for $P_v$:
\beqn\la{5.1}
P_v:&=&
-\Re\int\limits_{\R^2}
\dot\psi_v \overline{\nabla\psi_v}\,dx\nonumber\\
&=&
-\Re
\int\limits_{\R^2}
(-\gamma v(\nabla_1 a)(y)-i\omega\gamma a(y))
\Big(
\overline{\gamma (\nabla_1 a)(y)+i\omega\gamma va(y)},
\overline{\nabla_2 a(y)}
\Big)\frac{1}{\gamma}\,dy.
\eeqn
First,
\beqn\la{5.2}
&&\Re\int\limits_{\R^2}
(\nabla_1 a) (y)\overline{\nabla_2 a(y)}\,dy
\nonumber\\
&=&\Re
\int\limits_{0}^{+\infty}r\,dr
\int\limits_{0}^{2\pi}
\Big(R'(r)\cos\varphi
-ik\ds \frac{R(r)}{r}\sin k\varphi\Big)
\Big(R'(r)\sin k\varphi  -
ik\ds \frac{R(r)}{r}\cos k\varphi\Big)\,d\varphi
\nonumber\\
&=&\Re\int\limits_{0}^{+\infty}
\left[{R'}^2(r)-k^2\frac{R^2(r)}{r^2}\right]r\,dr
\int\limits_{0}^{2\pi}
\frac{\sin 2k\varphi}{2}\,d\varphi=0.
\eeqn
Similarly,
\beqn\la{5.3}
&&\Re\int\limits_{\R^2}
a (y)\overline{\nabla_2 a(y)}\,dy
=\Re\int\limits_{0}^{+\infty}r\,dr
\int\limits_{0}^{2\pi}
R(r)e^{ik\varphi}
\Big(R'(r)\sin k\varphi  -
ik\ds \frac{R(r)}{r}\cos k\varphi\Big)\,d\varphi
\nonumber\\
&=&\Re\int\limits_{0}^{+\infty}
\left[R(r)R'(r)+k\frac{R^2(r)}{r}
\right]r\,dr
\int\limits_{0}^{2\pi}
\frac{\sin 2k\varphi}{2}\,d\varphi=0.
\eeqn
Hence, relations (\ref{5.1})-(\ref{5.3}) imply
\beqn\la{5.4}
P_v:&=&-\Re\int\limits_{\R^2}
(-\gamma v(\nabla_1 a)(y)-i\omega\gamma a(y))
\Big(
\overline{\gamma (\nabla_1 a)(y)+i\omega\gamma va(y)},0
\Big)\frac{1}{\gamma}\,dy\nonumber\\
&=&
\Big(\gamma v\int\limits_{\R^2}
(|\nabla_1 a(y)|^2+\omega^2 a^2(y))
\,dy,0\Big)
=\Big(\gamma v 2(I_1+\omega^2 I_0),0\Big).
\eeqn
as in (\ref{6}). Further, relation (\ref{7})
and the equality $I_1=I_2$, which is already proved,
 imply the validity of (\ref{A}) for $P_v$.
\hfill$\Box$
\medskip\\
{\bf Remark.}
The existence of  solitary waves 
 $\psi_0(x,t)=R(r)e^{ik\phi}e^{-i\omega t}$ with $k\ne 0$
was not proved in \ci{BL}.
However, in this case,  equation (\re{a}) reduces to an
 ordinary differential equation
(as for radial solutions). Therefore,
 it is natural to think that  existence can be proved 
by modifing  the  method in \ci{BL}.


 \end{document}